\documentclass[fleqn,usenatbib]{mnras}

\usepackage{newtxtext,newtxmath}

\usepackage[T1]{fontenc}

\DeclareRobustCommand{\VAN}[3]{#2}
\let\VANthebibliography\thebibliography
\def\thebibliography{\DeclareRobustCommand{\VAN}[3]{##3}\VANthebibliography}

\def\draftversion{2} 
\usepackage{latexsym,graphicx,natbib,times,amsmath,xspace,ifthen}

\ifthenelse{ \draftversion > 0 }{
} {
  
}

\ifthenelse{\equal{\draftversion}{1}}{
  \usepackage{xcolor}
  \newcommand{\sep}[1]{\par\begin{color}[rgb]{0,0.4,0}\begin{center}\hrule\end{center}\end{color}\par} 
  \newcommand{\todo}[1]{\begin{color}{red}\ \ifthenelse{\equal{#1}{}} {$\bullet\bullet\bullet$} {$\bullet$\ #1 $\bullet$}\end{color}} 
  \newcommand{\idea}[1]{\begin{color}[rgb]{0,0.4,0}\textit{#1}\end{color}} 
  \newcommand{\sk}[1]{\begin{color}[rgb]{0.6,0,0.6}#1\end{color}} 
  \newcommand{\toc}{\par\begin{color}[rgb]{0.6,0,0.6}\begin{center}\hrule\vspace{0.5mm}\begingroup\small\let\cleardoublepage\relax\let\clearpage\relax\mytoc\endgroup\vspace{0.5mm}\hrule\end{center}\end{color}\par} 

  }{
  \newsavebox{\trashcan}

  \newcommand{\sep}[1]{}
  \newcommand{\todo}[1]{}
  \newcommand{\idea}[1]{}
  \newcommand{\sk}[1]{}
  \newcommand{\toc}{}

  }
 \makeatletter\newcommand\mytoc{\@starttoc{toc}}\makeatother 
\long\def\symbolfootnote[#1]#2{\begingroup%
\def\thefootnote{\fnsymbol{footnote}}\footnote[#1]{#2}\endgroup} 

\newcommand{\bb}[1]{\ifmmode \mbox{\boldmath $ #1$} \else  \mbox{\boldmath $#1$} \fi}

\newcommand{\U}[1]{\ensuremath{\mathrm{~#1}}}     

\newcommand{\yr}{\U{yr}}

\newcommand{\kpc}{\U{kpc}}
          
\newcommand{\Msun}{\U{M}_{\odot}}   \newcommand{\msun}{\Msun}
\newcommand{\Msunyr}{\Msun\yr^{-1}} 
\newcommand{\Zsun}{\U{Z}_{\odot}}   
\newcommand{\cc}{\U{cm^{-3}}}

\newcommand{\erg}{\U{erg}}

\newcommand{\feh}{\ensuremath{[\mathrm{Fe/H}]}\xspace}       

\newcommand{\ramses}{{\small RAMSES}\xspace}

\newcommand{\magi}{{\small MAGI}\xspace}

\newcommand{\strasbourg}{Observatoire Astronomique de Strasbourg, Universit\'e de Strasbourg, CNRS UMR 7550, 11 rue de l’Universit\'e, F-67000 Strasbourg, France}

\newcommand{\jerusalem}{Racah Institute of Physics, The Hebrew University, Jerusalem 91904, Israel}

\usepackage{graphicx}	
\usepackage{amsmath}	
\usepackage{gensymb}

\newcommand{\orcid}[1]{\href{https://orcid.org/#1}{\includegraphics[height=\fontcharht\font`\B]{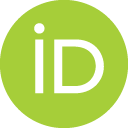}}}

\title[Star formation in a shell galaxy]{From starburst to quenching: merger-driven evolution of the \\ star formation regimes in a shell galaxy}

\author[Petersson et al.]{
Jonathan~Petersson$^{1, 2}$\thanks{E-mail: jonathan.petersson@epfl.ch} \orcid{0000-0001-7248-3898},
Florent~Renaud$^1$ \orcid{0000-0001-5073-2267}, 
Oscar~Agertz$^1$ \orcid{0000-0002-4287-1088}, 
Avishai~Dekel$^3$ \orcid{0000-0003-4174-0374}, 
Pierre-Alain~Duc$^4$ \orcid{0000-0003-3343-6284}
\\
$^1$Lund Observatory, Department of Astronomy and Theoretical Physics, Lund University, Box 43, SE-221 00 Lund, Sweden \\
$^2$Institute of Physics, Laboratory for Galaxy Evolution and Spectral Modelling, EPFL, Observatoire de Sauverny, Chemin Pegais 51, 1290 Versoix, Switzerland \\
$^3$\jerusalem \\
$^4$\strasbourg
}

\date{Accepted for publication by MNRAS}

\pubyear{2022}

\begin{document}
\label{firstpage}
\pagerange{\pageref{firstpage}--\pageref{lastpage}}
\maketitle

\begin{abstract}
Shell galaxies make a class of tidally distorted galaxies, characterised by wide concentric arc(s), extending out to large galactocentric distances with sharp outer edges. Recent observations of young massive star clusters in the prominent outer shell of NGC 474 suggest that such systems host extreme conditions of star formation. In this paper, we present a hydrodynamic simulation of a galaxy merger and its transformation into a shell galaxy. We analyse how the star formation activity evolves with time, location-wise within the system, and what are the physical conditions for star formation. During the interaction, an excess of dense gas appears, triggering a starburst, i.e. an enhanced star formation rate and a reduced depletion time. Star formation coincides with regions of high molecular gas fraction, such as the galactic nucleus, spiral arms, and occasionally the tidal debris during the early stages of the merger. Tidal interactions scatter stars into a stellar spheroid, while the gas cools down and reforms a disc. The morphological transformation after coalescence stabilises the gas and thus quenches star formation, without the need for feedback from an active galactic nucleus. This evolution shows similarities with a compaction scenario for compact quenched spheroids at high-redshift, yet without a long red nugget phase. Shells appear after coalescence, during the quenched phase, implying that they do not host the conditions necessary for in situ star formation. The results suggest that shell-forming mergers might be part of the process of turning blue late-type galaxies into red and dead early-types. 
\end{abstract}

\begin{keywords}
galaxies: interactions -- galaxies: starburst -- galaxies: star formation -- methods: numerical
\end{keywords}


\section{Introduction}
In the current cosmological paradigm, galaxy interactions and mergers are natural recurring events \citep{White1978}, often associated with starbursts and enhanced star formation activity \citep{Sanders1996, Canalizo2001, Knapen2015}. With the development of more powerful supercomputers throughout the last few decades, numerical simulations have provided a theoretical framework to understand the evolution and underlying physics of the star formation activity in a wide diversity of interacting galaxies and mergers \citep{Hernquist1989, Barnes1991, Mihos1996, DiMatteo2007}. Non-cosmological merger simulations have proven to be especially useful for this, because they allow to recreate specific tidally distorted galaxies \citep{Toomre1972}, and enable the analysis of the star formation activity at high spatial resolution \citep{Karl2010, Teyssier2010, Renaud2014, Moreno2015, Renaud2018}.

One particular class of tidally distorted galaxies are shell galaxies \citep{Arp1966}, characterised by wide concentric arc(s), extending out to large galactocentric distances with sharp outer edges. Several formation mechanisms have been proposed for shells in the last decades (for a review, see e.g. \citealt{Ebrova2013, Pop2018}). The most widely accepted invokes a merger event. With the help of numerical simulations, it has been inferred that shells can form via near-radial (i.e. a small impact parameter) intermediate to major mergers (i.e. a mass ratio $\gtrsim$1:10). In this scenario, shells consist of tidally stripped stars from an accreted satellite galaxy, accumulating at the apocentres of their orbits \citep{Quinn1984, Hernquist1992, Karademir2019, Pop2018}. This tidal origin is shared with bridges and tails, but with the latter tending to be more favourably produced in less eccentric mergers \citep{Amorisco2015, Hendel2015}.

Observations have shown that shells are more frequent around red early-type galaxies than blue late-types \citep{Atkinson2013}, and that $\sim$20\% of elliptical and lenticular galaxies exhibit one or more shells \citep{Malin1983, Tal2009, Duc2015, Bilek2020}. However, since the visibility of the shells may depend on their orientation and age \citep[see e.g.][]{Krajnovic2011}, such estimates from observations suffer from uncertainties. Nevertheless, shells can remain visible in the galaxies' stellar haloes several Gyr after the merger event itself \citep{Mihos2005, MartinezDelgado2010, Duc2015, Mancillas2019b}, and therefore serve as a powerful tool to help reconstruct assembly histories of galaxies \citep{Romanowsky2012, Foster2014, Longobardi2015}. 

Due to its numerous prominent shells, NGC 474 is a prototypical system of a shell galaxy, often put in the spotlight when studying the formation of shells and shell galaxies. The recent simulation of \citet{Bilek2022} confirms that a shell morphology similar to that of NGC 474 could be explained by a radial interaction. However, besides constraints on its formation history, recent works have also reported some oddities. The outer shell of NGC 474 seems to host several globular clusters \citep{Lim2017}, a result spectroscopically confirmed by two independent teams \citep{Alabi2020, Fensch2020}. Among these massive star clusters, two are relatively young with respect to the age of the overall stellar population making the host shell \citep{Fensch2020}. Instead, the ages and metallicities of the two clusters match that of the much younger stellar population in the nucleus of NGC 474, far away from the tidal debris. Pinpointing the origin of these clusters, and assessing the star formation activity in shell galaxies in general, calls for numerical simulations that can reproduce the physical conditions for star and cluster formation in these type of systems. However, the simulation of \citet{Bilek2022} does not include gas dynamics. New simulations are therefore required to address the question of the origin(s) of clusters in outer shells, and more generally when and where star formation occurs in shell galaxies.

To understand how the merger-driven star formation activity in a shell galaxy evolves with time and within the system, we present here the analysis of a new non-cosmological hydrodynamic simulation of a shell-forming merger. The paper is organised as follows. In Section~\ref{sect:method}, we present the numerical set-up and sub-grid models of our simulation. Section~\ref{sect:results} focuses on the global and spatially resolved star formation activity, and how it relates to the shells of the merger. In Section~\ref{sect:discussion}, we discuss the implications of our results on the formation history of NGC 474, and on shell galaxies in general. Section~\ref{sect:conclusions} presents the conclusions of our study. 

\section{Method}
\label{sect:method}
\subsection{Numerical set-up}
We run a \textit{N}-body + hydrodynamical simulation of two merging disc galaxies and their transformation into a shell galaxy, using the adaptive mesh refinement (AMR) code \ramses \citep{Teyssier2002}. Dark matter (DM) and stars are treated as collisionless particles, whereas the fluid dynamics of the gas is computed on the AMR grid, assuming ideal mono-atomic gas with an adiabatic index $\gamma=5/3$. The grid follows a quasi-Lagrangian refinement strategy (see \citealt{Renaud2015}), down to a maximum physical resolution of $\sim$12 pc. 

The two galaxy models are generated in isolation using the \magi code \citep{Miki2018}, both with a dark matter halo, a central bulge, a stellar and a gaseous disc. The density profiles between the two models are identical, and the initial parameters are provided in Table~\ref{tab:initial_conditions}. The models are rendered with a total of 2 million particles (excluding the gas which is rendered on the AMR grid), distributed accordingly to each galaxies' respective components, giving maximum mass resolutions of $2.4\times10^4 \Msun$ and $9.6\times 10^5\Msun$ for the stellar and dark matter particles respectively. The initial total mass ratio between the two galaxies is 1:2\footnote{With a short parameter survey of low-resolution merger simulations (not shown here), we find that this mass ratio is the most favourable for shell formation.}. The most massive galaxy is henceforward referred to as the ``main galaxy'', and the less massive galaxy as the ``satellite''. 

In the merger simulation, the satellite has an initial inclination angle of $10\degree$ with respect to the plane of the main galaxy, to avoid the special coplanar case. The main galaxy is initially placed at the centre of a $200 \kpc$ cube (filled with an ambient gaseous medium of $10^{-5} \cc$) and the satellite a few tens of kpc away with a set of orbital parameters to give it a near-radial infall trajectory towards the main galaxy. The initial separation between the two galaxies is set large enough to allow both galaxies to relax from the artificial initial conditions and reach a steady star formation activity before the encounter, yet sufficiently short to limit the computational cost, and the premature consumption of the gas reservoirs. The orbital parameters of the encounter are given in Table~\ref{tab:orbital_parameters}.

\subsection{Sub-grid models}
\label{sect:sub-grid_models}
Gas cooling is metallicity-dependent and determined using tabulated values from the cooling functions by \citet{Sutherland1993} for gas temperatures in the range of $10^{4 - 8.5}$ K, and from \citet{Rosen1995} for temperatures $<$ $10^4$ K. Heating from ultraviolet background radiation is modelled according to \citet{Haardt1996}. The initial metallicity of the gas reservoirs of the main galaxy and the satellite are $1\Zsun$ and $0.3\Zsun$ respectively. 

Star formation and feedback physics follows that of \citet{Agertz2013, Agertz2015}. In short, star formation occurs in individual cells according to the star formation law:

\begin{equation}
	\rho_\mathrm{SFR} =  \epsilon_\mathrm{ff} \frac{\rho_\mathrm{g}}{t_\mathrm{ff}} \quad \mathrm{when} \quad \rho_\mathrm{g} > \rho_\mathrm{SF},
	\label{eq:StarFormationLaw}
\end{equation}

\noindent where $\rho_\mathrm{SFR}$ is the star formation rate (SFR) density, $\rho_\mathrm{g}$ is the gas density, $\epsilon_\mathrm{ff}=0.1$ is the star formation efficiency per free-fall time $t_\mathrm{ff}=\sqrt{(3\pi)/(32G\rho)}$, and $\rho_\mathrm{SF}=100$ cm$^{-1}$ is the star formation density threshold. Star particles of $10^4 \msun$ are generated via a Poisson sampling process, and each particle is treated as a single-age stellar population with a \citet{Chabrier2003} initial mass function (IMF). Star formation is regulated via the injection of energy, momentum, mass and heavy metals into surrounding gas through stellar feedback processes, including types II and Ia supernovae (SNe), stellar winds, and radiation pressure. To avoid numerical overcooling in cases where the Sedov-Taylor phase of individual SNe is not resolved, we implement the model suggested by \citet{Kim2015}, where the cooling radius must stretch over at least three grid cells to capture the momentum injection, with an energy of $10^{51} \erg$ per SNe. 

\section{Results}
\label{sect:results}
\begin{figure*}
	\includegraphics[width=2\columnwidth]{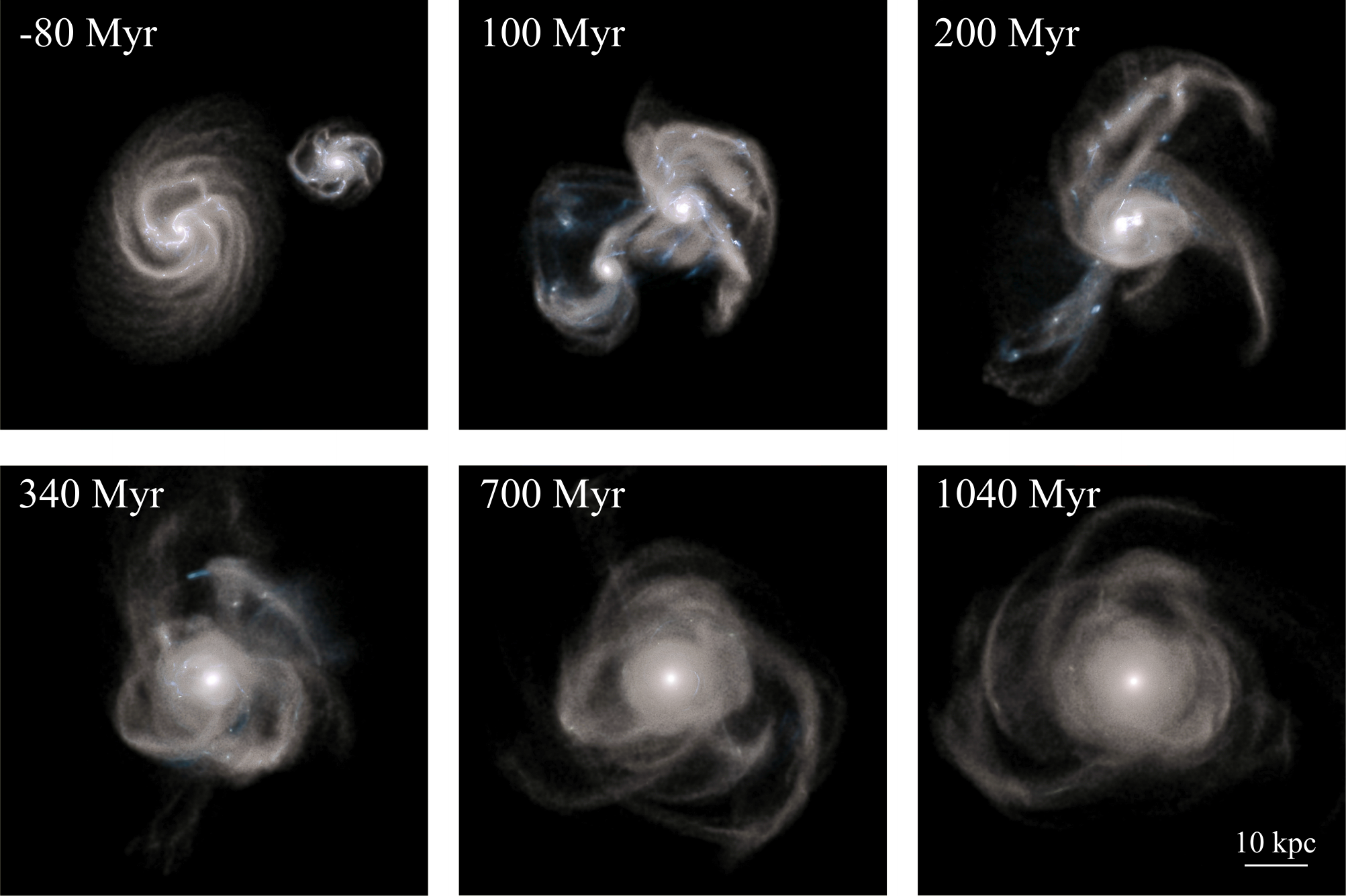}
	\caption{Face-on mock composite images (using the u, g and i filters of the MegaCAM instrument on the Canada-France-Hawaii Telescope) of our simulation (see Appendix \ref{appendix:mock} for details on mock observations). The first stellar shell appears at $t = 340$ Myr in the top right part of the galaxy. Another pair of shells can be seen at $t = 700$ Myr and $t = 1040$ Myr, in the bottom right and middle left part of the galaxy respectively.}
	\label{fig:composite}	
\end{figure*}

Fig.~\ref{fig:composite} shows face-on mock observations of the simulated system before, during, and after the merger event ($t=0$ is defined to be at the first pericentre passage). The first distinct stellar shell appears at $t = 340$ Myr and can be seen in the top right part of the galaxy. Stellar shells continue to form for about 1 Gyr, but become increasingly difficult to detect over time as their dissolution causes a lowering of their surface brightness. These stellar shells do not exhibit a gaseous counterpart, as shown in more detail in Section~\ref{sect:shellsandgas}. The interaction also disrupts the main galaxy, with a morphological transformation of its disc into a spheroid. 

\subsection{Global star formation history}
\label{sect:globalSFH}

\begin{figure}
	\includegraphics[width=\columnwidth]{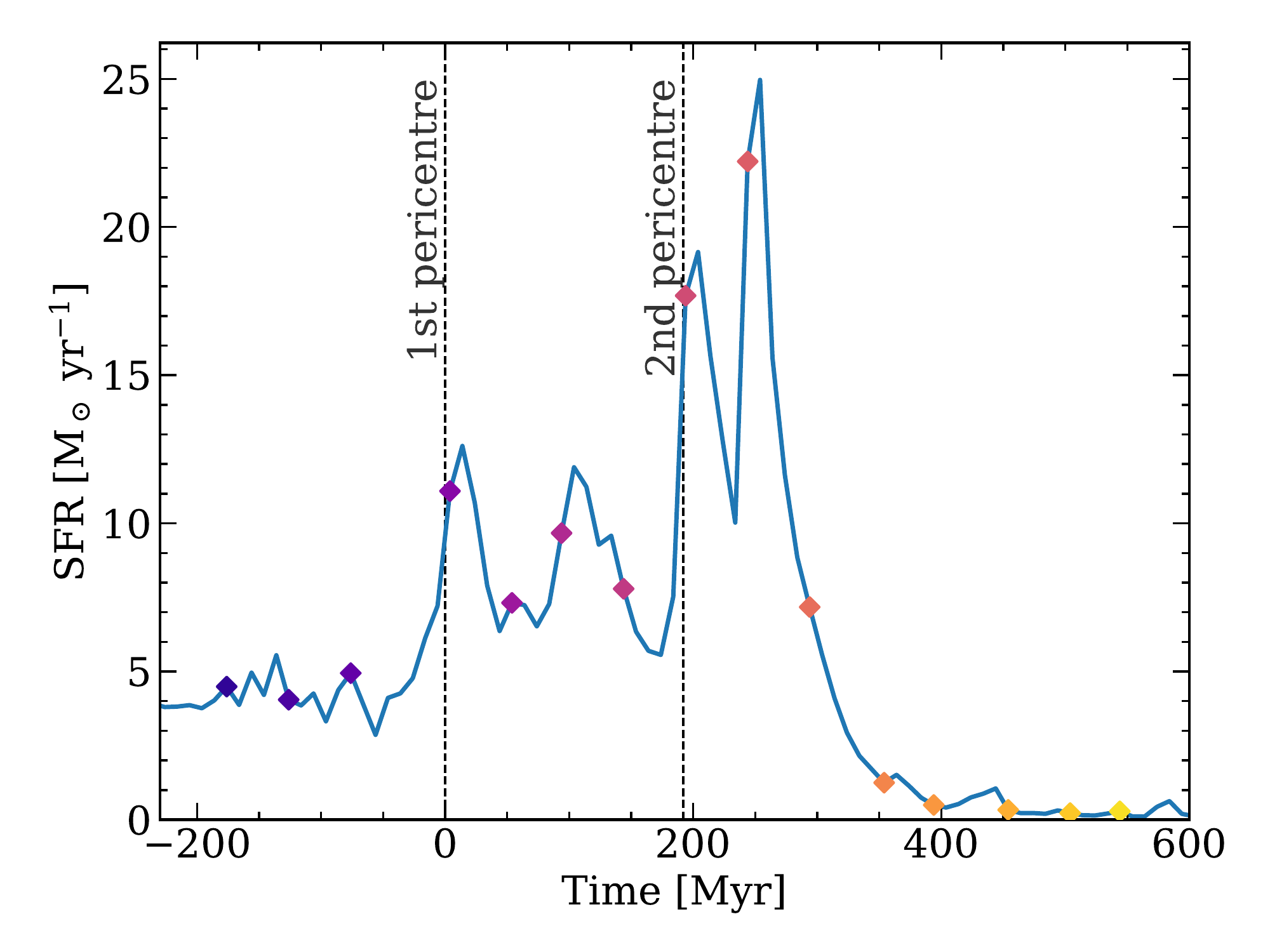}
	\caption{Evolution of the SFR in the simulation for a time period covering all epochs of star formation. As expected in wet mergers, the pericentre passages are associated with significant enhancements of the star formation activity. Coloured symbols mark snapshots considered in the rest of the analysis.}
	\label{fig:SFR}
\end{figure}

In Fig.~\ref{fig:SFR}, we show the evolution of the SFR in the simulation, from  230 Myr before to 600 Myr after the first pericentre passage, i.e. sufficiently to cover all the epochs of star formation. Before the first pericentre passage ($t<0 $), the pre-interaction phase yields a relatively constant SFR of about $4 \Msunyr$. As we approach the first pericentre passage, the SFR increases and the system goes into the interaction phase with several short-lived ($\sim$30 Myr) peaks. The SFR increases even further as we get close to the second pericentre passage. After $t = 250$ Myr, it rapidly decreases and goes well below its pre-interaction value. The SFR never recovers, and this quenching phase continues throughout the rest of the simulation (i.e. up to 4 Gyr after the first pericentre passage). 

\begin{figure}
	\includegraphics[width=\columnwidth]{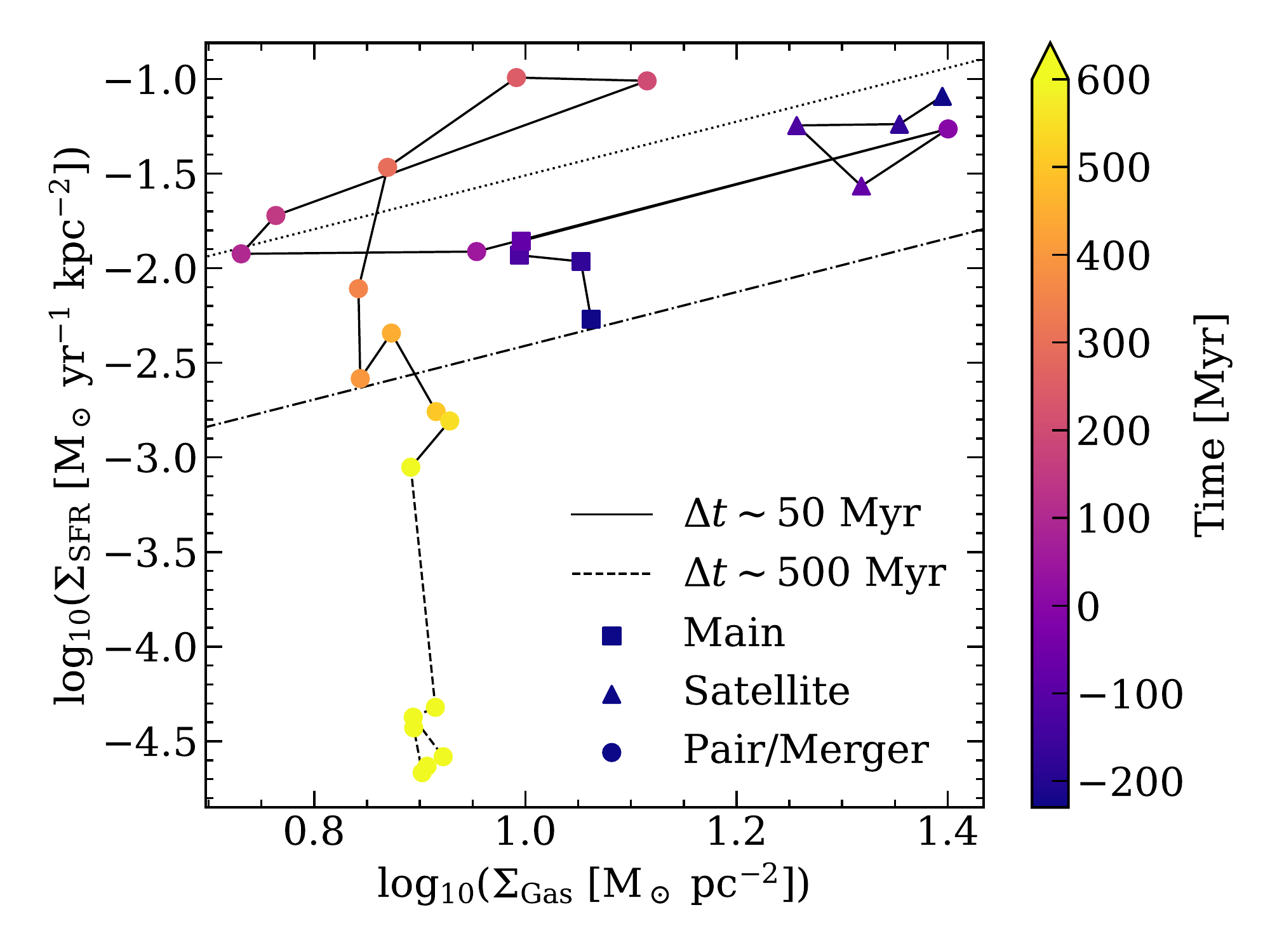}
	\caption{Time evolution of surface density of gas and SFR within two times the half-mass stellar radius, initially every $\sim$50 Myr, and every  $\sim$500 Myr after $t=600$ Myr. The upper dotted line is a fit to observations of local (ultra) luminous infrared galaxies (ULIRGs) and submillimeter galaxies (SMGs), while the lower dash-dotted line is a fit to local spirals and $z = 1.5$ BzK galaxies \citep{Daddi2010}.}
	\label{fig:KSEvol}
\end{figure}

To understand what physical processes are causing these distinct epochs of star formation in Fig.~\ref{fig:SFR}, we explore the time evolution of the SFR with respect to the gas density in a Kennicutt-Schmidt (KS) diagram. Figure~\ref{fig:KSEvol} shows the time evolution of the surface density of gas and SFR within two times the half-mass stellar radius, initially every $\sim$50 Myr (as indicated by the symbols in Fig.~\ref{fig:SFR}), and every $\sim$500 Myr after $t=600$ Myr. In the pre-interaction phase, the two galaxies share similar behaviours in the KS-plane, and do not experience any drastic changes in either surface densities. During the interaction phase, the system starts with a high surface density of gas, which then rapidly decreases while the surface density of SFR increases. This reduces the depletion time of the system (to a minimum of the order of 100 Myr), marking the onset of the starburst regime. The quenching phase ($t>250$ Myr) starts with a violent decrease of the SFR surface density, while that of the gas remains close to constant. This leads to an increase in depletion time, with a maximum reaching $\sim$100 Gyr. 

\begin{figure}
	\includegraphics[width=\columnwidth]{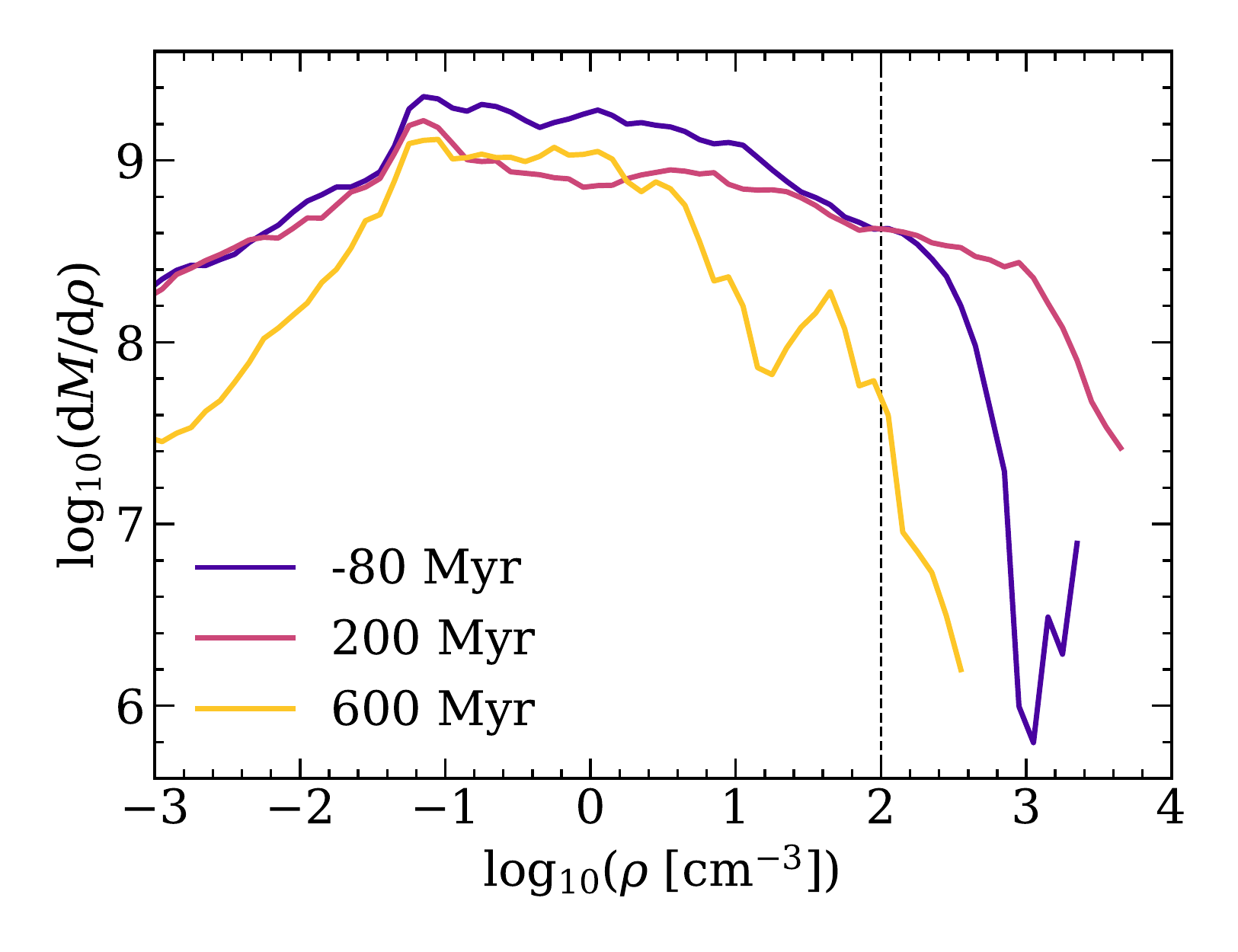}
	\caption{Mass-weighted probability distribution functions of the gas density at three different snapshots, representing the different epochs of star formation in our simulation. This includes the pre-interaction phase ($t = -80$ Myr), the interaction/starburst phase ($t = 200$ Myr) and the quenching phase ($t = 600$ Myr). The dashed line marks the star formation threshold at $\rho_\mathrm{SF} = 100$ cm$^{-1}$.}
	\label{fig:PDF}
\end{figure}

\begin{figure}
	\includegraphics[width=\columnwidth]{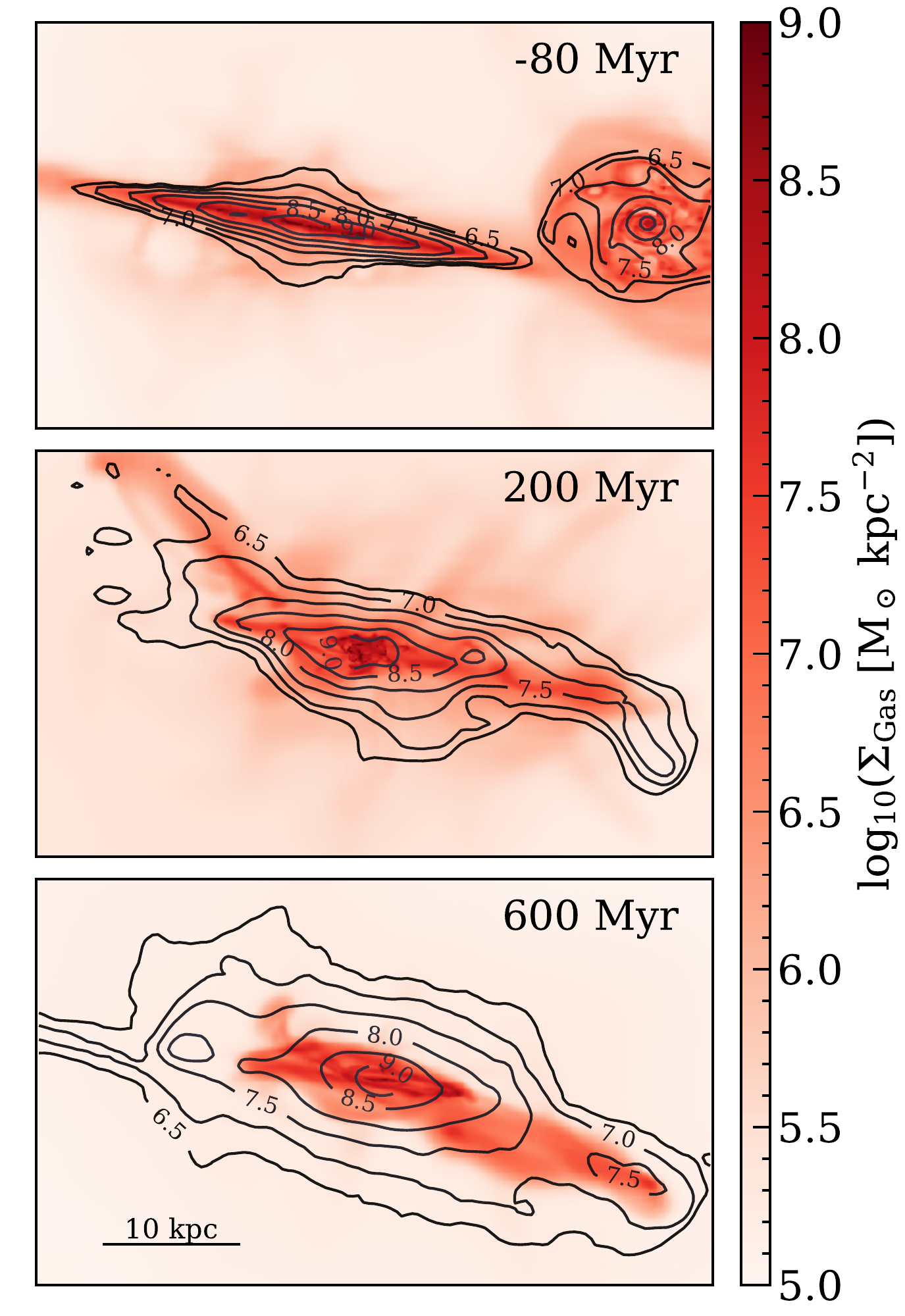}
	\caption{Edge-on surface density of gas together with contours of the stellar surface density (with values of  $\log_{10}(\Sigma_\star [\mathrm{M}_\odot \ \mathrm{kpc}^{-2}])$ shown) for the same three snapshots as in Fig.~\ref{fig:PDF}, representing the different epochs of star formation. Tidal interactions during the merger puffs up the stellar component of the main galaxy into a spheroid, which weakens the gravitational force on the gas and makes it stable against collapse.}
	\label{fig:MQ}
\end{figure}

The fact that the surface density of gas remains close to constant throughout both the starburst and quenching phase in Fig.~\ref{fig:KSEvol} suggests that internal redistributions of the gas density are responsible for the different phases of star formation. In Fig.~\ref{fig:PDF}, we show the mass-weighted probability density functions of the gas density at three different snapshots, illustrative of the three epochs of star formation identified in Fig.~\ref{fig:SFR} and~\ref{fig:KSEvol}. For the starburst phase ($t=200$ Myr), we see an excess of high density gas with respect to the pre-interaction. In galaxy interactions, several physical mechanisms can lead to an increase in mass of dense gas and/or a decrease of depletion time, and by that trigger a starburst: nuclear inflows induced by tidal torques \citep{Barnes1991}, shocks between gas reservoirs when dense regions of the ISM overlap \citep{Jog1992}, and tidal and turbulent compression of gas \citep{Renaud2014}. The relative role played by each of these mechanisms is highly space- and time-dependent and remains difficult to quantify. While nuclear inflows are restricted, by definition, in the galactic centres, the other processes can also be active there, and disentangling them is a tedious task (see an example in \citealt{Renaud2019}). Yet, only tidal and turbulence compression can explain off-centre starbursts in non-overlapping volumes (i.e. where shocks cannot be found).

In our simulation, the gas mass fraction in compressive tides increases by a factor of $\sim$1.3 at the times of the pericentre passages, confirming the ubiquity of tidal compression in mergers, including shell-forming configurations \citep{Renaud2009}, although at lower levels than those found in simulations with higher peaks of SFR (e.g. in Antennae-like systems, and collisional rings, see \citealt{Renaud2014, Renaud2018} respectively). Tidally compressed volumes are found near star forming regions, such as the galactic centre, but also in off-centered regions (see Section~\ref{sect:resolvedSFA}). The SFR peaks about 30 Myr after the onset of tidal compression, thus also after the pericentre passages themselves (recall Fig.~\ref{fig:SFR} showing that the SFR peaks 10--30 Myr after the pericentre passages), as also found by \citet{Renaud2014}, \citet{Moreno2019}, and \citet{Li2022}. This delay can be explained by the time needed for tidal compression to be converted into compressive turbulence, and for this energy to cascade down to the scales of star formation \citep{Renaud2019}.

\begin{figure*}
	\includegraphics[width=2\columnwidth]{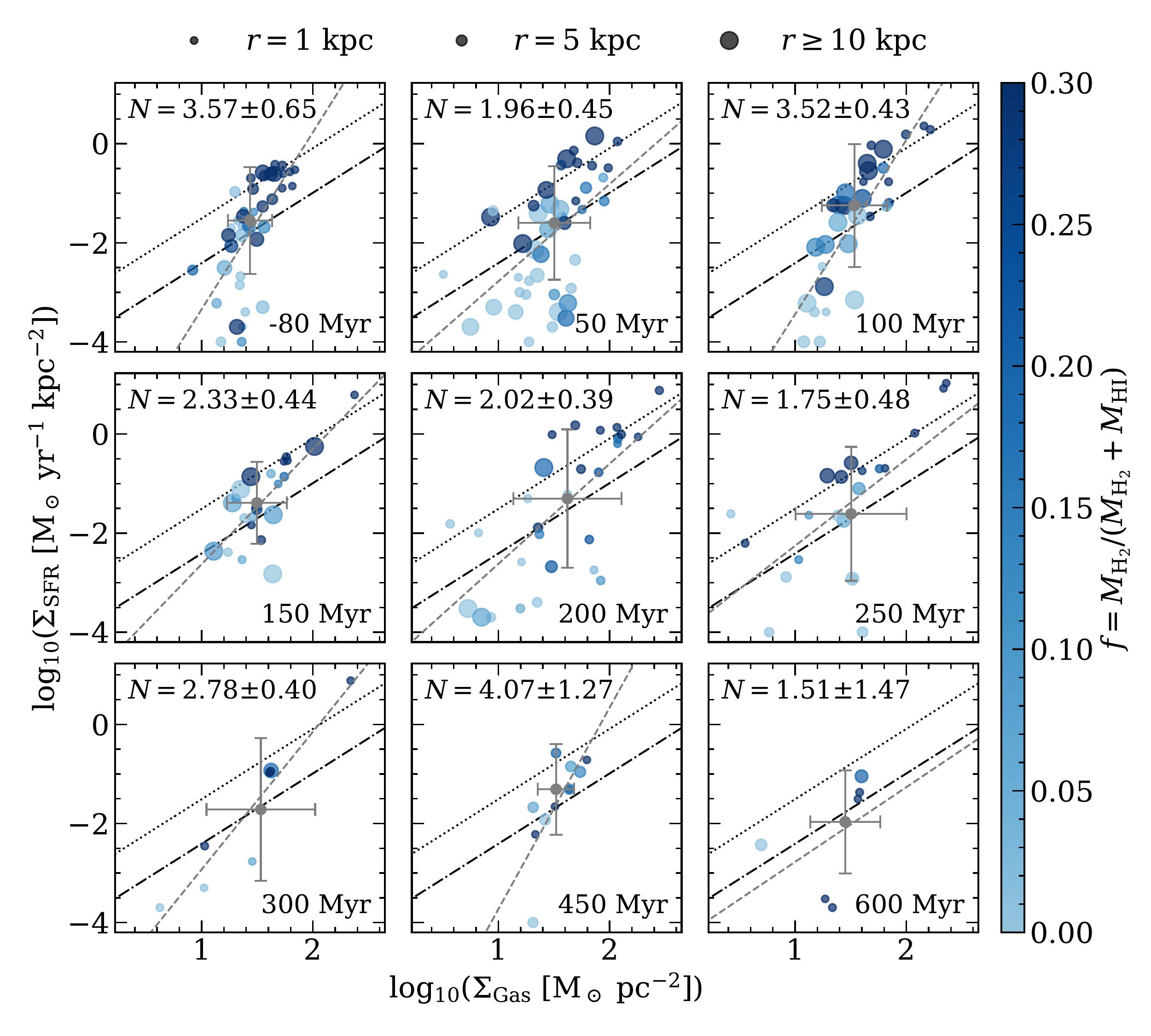}
	\caption{Spatially resolved KS-diagrams of our simulation at nine different snapshots (chosen to include all epochs of star formation from Section~\ref{sect:globalSFH}), accounting for all the gas (i.e. molecular, atomic and ionised gas). Each point in each diagram is a star-forming region (considering stars younger than 10 Myr) on a 1 kpc scale, colour-coded and scaled in size according to its molecular gas fraction and distance from the centre of (stellar) mass of the system respectively. At $t=-80$, only star-forming regions from the main galaxy are included. For each diagram, a median value of the surface densities is calculated (grey dot), and a power-law is fitted (grey dashed line), according to the star formation law $\Sigma_\mathrm{SFR} \propto \Sigma_\mathrm{Gas}^N$ \citep{Schmidt1959, Kennicutt1998}.}
	\label{fig:KSRelations}
\end{figure*}

In the quenching phase ($t = 600$ Myr), Fig.~\ref{fig:PDF} shows a deficit of high density gas. Examples of physical processes that can lead to less amounts of high density gas includes the consumption of all the gas by the starburst, and the removal of large amounts of the gas through tidal ejections, feedback processes (outflows) and/or ram pressure stripping. In the time period $-80 \ \mathrm{Myr} < t < 600 \ \mathrm{Myr}$, the mass of gas within two times the half-mass stellar radius decreases by $2.49\times 10^9 \Msun$, while the stellar mass increases by $2.55 \times 10^9 \Msun$ over the same period. This shows that star formation is the main contributor to the decrease of the gas mass, however, as Fig.~\ref{fig:KSEvol} and~\ref{fig:PDF} suggest, it is not enough to empty the gas reservoir. The quenching phase is therefore not caused by a (hydro)dynamical removal of the gas reservoir, but rather by the dispersal of the dense phase into a more diffuse and not star-forming medium.

\begin{figure*}
	\includegraphics[width=2\columnwidth]{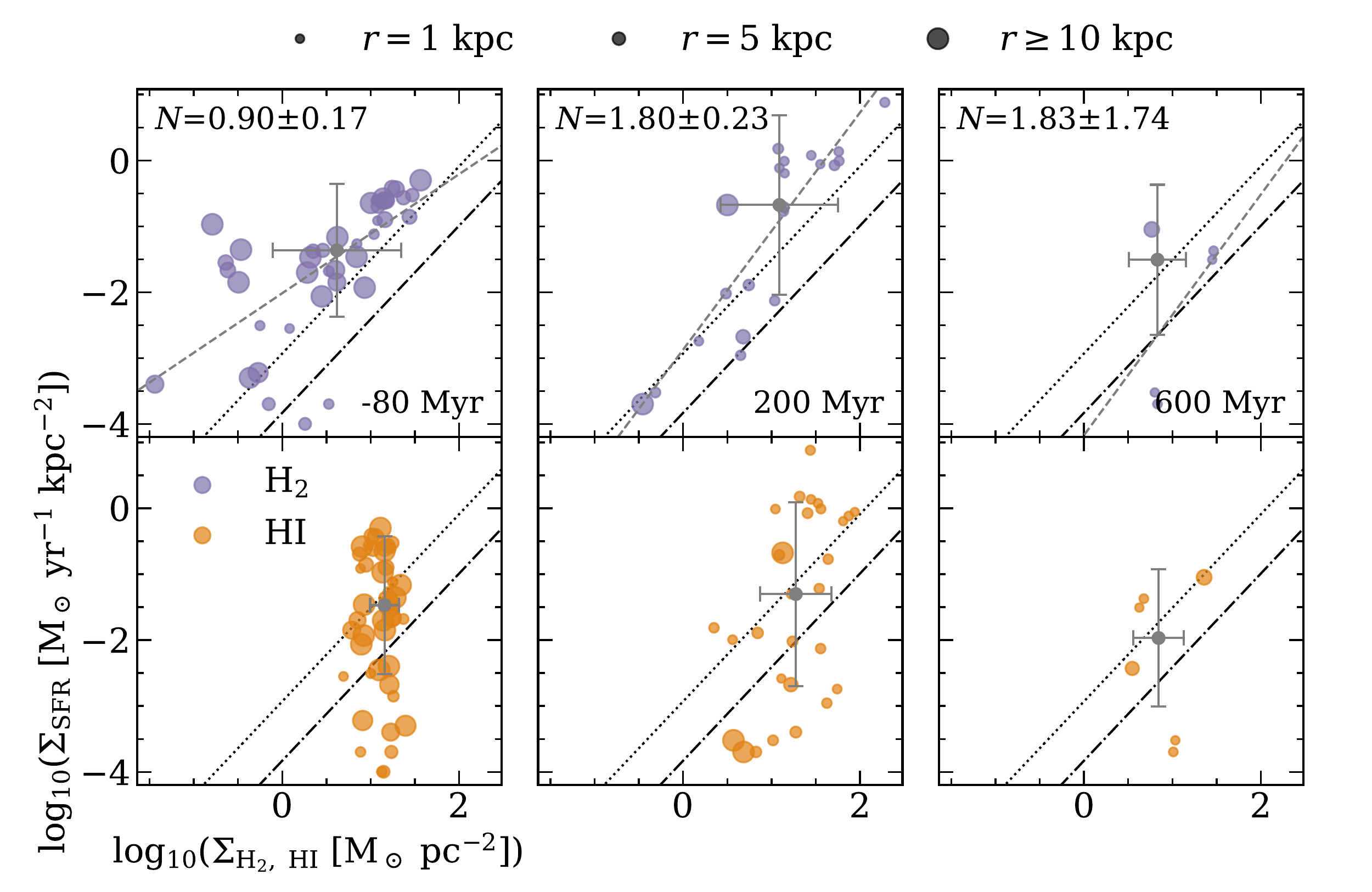}
	\caption{Similar set of KS-diagrams as in Fig.~\ref{fig:KSRelations}, but now separated into surface densities of H$_2$ (top) and HI (bottom), at snapshots representative for each epoch of star formation. No power-law is fitted for HI because no correlation between the surface density of SFR and HI can be distinguished.}
	\label{fig:KSRelationsMolecularAtomic}
\end{figure*}

Morphological quenching \citep{Martig2009} happens when the stellar component of a galaxy goes from being a disc to a spheroid. The gravitational potential then gets shallower, and the gas disc stabilises against fragmentation and gravitational collapse: the gas can no longer get dense enough to form stars. Fig.~\ref{fig:MQ} illustrates the growth of a stellar spheroid in our simulation by showing the edge-on surface density of gas, together with contours of the stellar surface density for the same three snapshots as in Fig.~\ref{fig:PDF}. The gas component also gets a more spheroidal-shape, but not to the same extent as the stellar component, probably because of its dissipative nature \citep[see e.g.][]{Athanassoula2016}. Using the same method as in \citet{Segovia2022} for determining whether the circular motion of stars ${v_\mathrm{rot, s}}$ dominate over the dispersion $\sigma_\mathrm{s}$, we estimate that $v_\mathrm{rot, s}/\sigma_\mathrm{s}\sim 5.0$ before ($t=-80$ Myr), and $v_\mathrm{rot, s}/\sigma_\mathrm{s}\sim 1.1$ after the merger ($t=600$ Myr), indicating that the stellar component ceases to be rotationally-supported. Morphological quenching is therefore very likely the main cause for the drop in the SFR after the interaction phase. 

\subsection{Resolved star formation activity}
\label{sect:resolvedSFA}
Figure~\ref{fig:KSRelations} shows a series of KS-diagrams at nine different snapshots in our simulation, where each point in each panel represents a gridded $1\times 1$ kpc scale region containing stars younger than 10 Myr, colour-coded by its molecular gas fraction inferred from the gas-phase diagram\footnote{The molecular gas fraction is defined as $f=M_{\mathrm{H}_2}/(M_{\mathrm{H}_2} + M_\mathrm{HI})$, where HI and H$_2$ respectively correspond to the cool ($T<8000$ K, $n<10$ cm$^{-3}$) and cold-dense ($T<300$ K, $n>10$ cm$^{-3}$) phases of the ISM, as in \citet{Moreno2019}}, and scaled in size according to its distance from the stellar centre of mass of the system. 

For each panel of Fig.~\ref{fig:KSRelations}, we calculate the median value of the 1 kpc regions (grey dot). This value does not significantly evolve, but instead remains within one standard deviation throughout all snapshots. The median alone does not reflect the evolution of the star formation activity (pre-interaction, starburst, quenching) highlighted in the previous section in the global analysis (Fig.~\ref{fig:KSEvol}). This implies that the median of the star forming regions is not a good representation of the galactic-scale star formation activity. The discrepancy originates from the dilution of the star formation signal when more and more non-star forming medium is accounted for in the unresolved analysis.

In Fig.~\ref{fig:KSRelations}, we fit a power-law (grey dashed line) of the form $\Sigma_\mathrm{SFR} \propto \Sigma_\mathrm{Gas}^N$ \citep{Schmidt1959, Kennicutt1998} in each diagram. The value of the power-law index $N$ varies strongly between snapshots, but with no general trend with time, nor with the epoch of star formation. However, the standard deviation of $N$ increases in late snapshots due to the reduction of the number of star-forming regions by quenching, which degrades the quality of the fits. However, despite strong variations, $N$ remains consistently higher than the classical value of 1.4 \citep{Schmidt1959, Kennicutt1998}. We also find a dependence of the molecular gas fraction with the surface densities, where star-forming regions with a higher molecular gas fraction yield shallower KS slopes, better in line with that reported by observations of molecular gas only \citep[e.g.][]{Bigiel2008}. However, contrary to \citet{Bigiel2008}, there is no visible dependence of $N$ with the galactocentric distance. This is likely due to the different mechanisms mentioned in Section~\ref{sect:globalSFH} like nuclear inflows, and off-centred starbursts due to shocks and/or tidal compression \citep{Renaud2019}. We suspect that the break in the KS-distribution relates to the mixing of HI and H$_2$, and is responsible for the high values of $N$ we find when considering all the gas. 

\begin{figure*}
	\includegraphics[width=2\columnwidth]{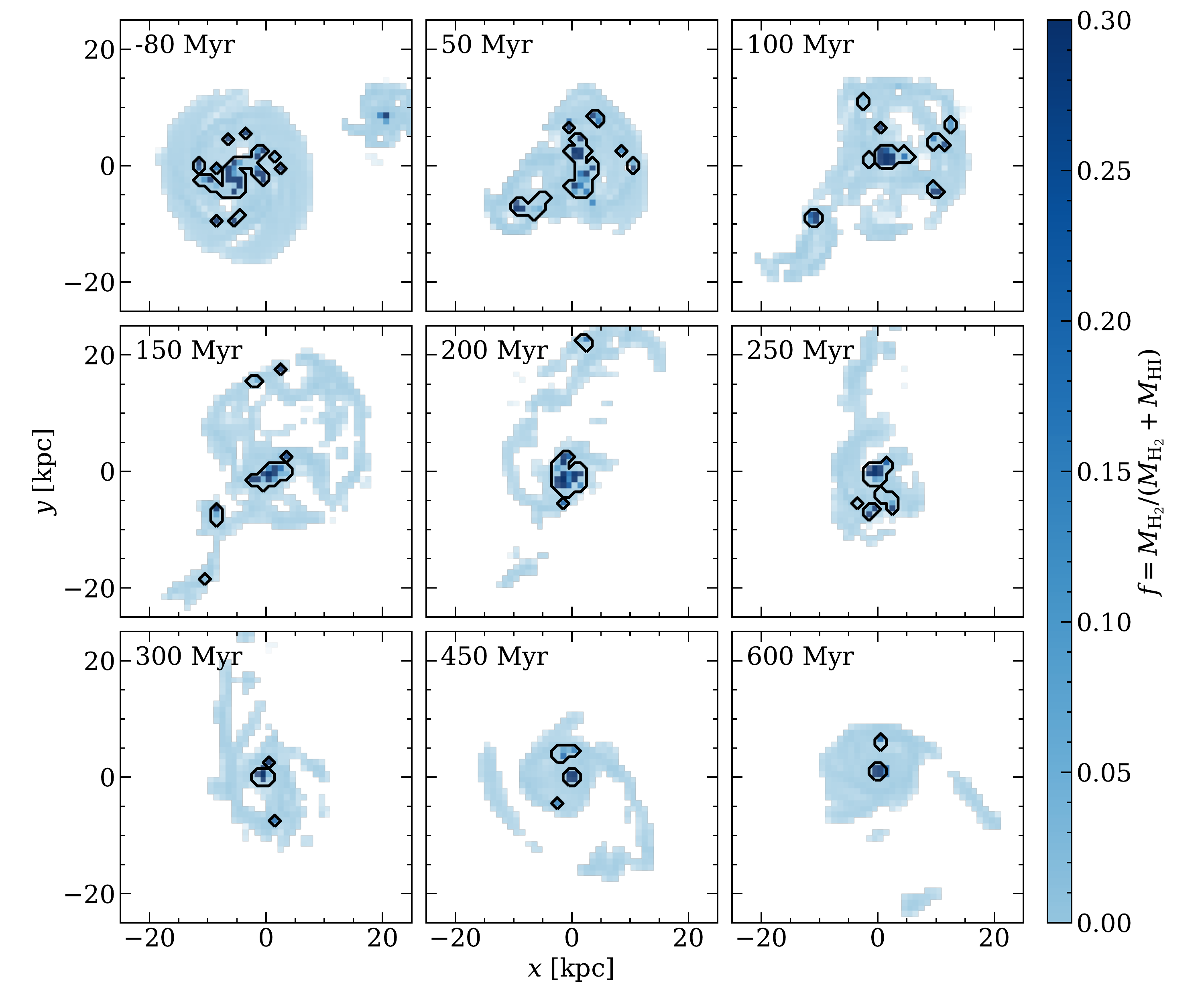}
	\caption{Face-on maps of the molecular gas fraction for the same snapshots as in Fig.~\ref{fig:KSRelations}. Each pixel (1 kpc scale) is mass-weighted in transparency, with more opaque pixels having a higher total gas mass. Black contours indicate star-forming regions (containing stars younger than 10 Myr). Star-forming regions from the satellite in the first panel at $t=-80$ Myr are not taken into account.}
	\label{fig:moleculargasfractionmaps}
\end{figure*}

The existence of a break in the KS-distribution at low gas surface densities is well known from previous research. By studying the star formation relation of nearby galaxies on sub-kpc scales, \citet{Bigiel2008} relates the break to the transition from an HI to an H$_2$-dominated ISM. However, the underlying physical origin of the break is still debated. \citet{Krumholz2009} proposes a theoretical model in which the break and the transition from HI to H$_2$ corresponds to self-shielding of hydrogen \citep[see also][]{Schaye2004}. However, the analytical model of \citet{Renaud2012} and simulations of \citet{Kraljic2014} suggest instead that the break is linked to the transition from subsonic to supersonic turbulence in the ISM. Without going into a deep analysis of the origin(s) of the break, we point out that the break exists in all stages of the merger considered here. It is however less pronounced during the peaks of SFR, when turbulence reaches its maximum level.

In Fig.~\ref{fig:KSRelationsMolecularAtomic}, we separate the total gas surface density into surface densities of HI and H$_2$, and examine how each relates to the surface density of SFR, in three selected snapshots representative of each phase of star formation. The surface density of H$_2$ shows a much tighter correlation with the surface density of SFR, but also a KS-relation with $N$ close to unity, as found empirically \citep{Bigiel2008}. The surface density of HI yield the opposite effect, i.e. no clear correlation with the surface density of SFR. Fig.~\ref{fig:KSRelationsMolecularAtomic} shows that the depletion time is close to constant for H$_2$-dominated star-forming regions (matching the $N\approx 1$ slope), while it varies by several dex in HI.

In contrast to isolated disc galaxies \citep{Bigiel2008}, no correlation between the molecular gas fraction or the depletion time with the galactocentric radius can be seen in Fig.~\ref{fig:KSRelations} or~\ref{fig:KSRelationsMolecularAtomic}. This is most likely due to the mixing of HI and H$_2$ induced by tidal interactions during the merger, as well as the trigger of off-centre starbursts. To investigate where star formation occurs, Fig.~\ref{fig:moleculargasfractionmaps} maps the distribution of molecular gas for each snapshot in Fig.~\ref{fig:KSRelations}, and identifies star-forming regions with black contours. The central kiloparsecs of the two galaxies and merger host high molecular gas fractions, corresponding to a major star formation activity. However, star formation also takes place in the spiral arms of the main galaxy before the encounter, and occasionally in the outskirts of the system during the early stages of the interaction. Star-forming regions at large galactocentric distances correspond to tidally compressed volumes, since tidal torques and shocks between gas reservoirs and/or clouds are not as effective here as they are close to the nucleus \citep{Renaud2014, Renaud2019}. As the overall star formation slows down ($t>250$ Myr), the off-centre activity fades, and only the innermost $\sim$1.5 kpc still contains star-forming molecular gas when the quenching phase starts.

\begin{figure*}
	\includegraphics[trim={1cm 0cm 1cm 0cm}, clip, width=2\columnwidth]{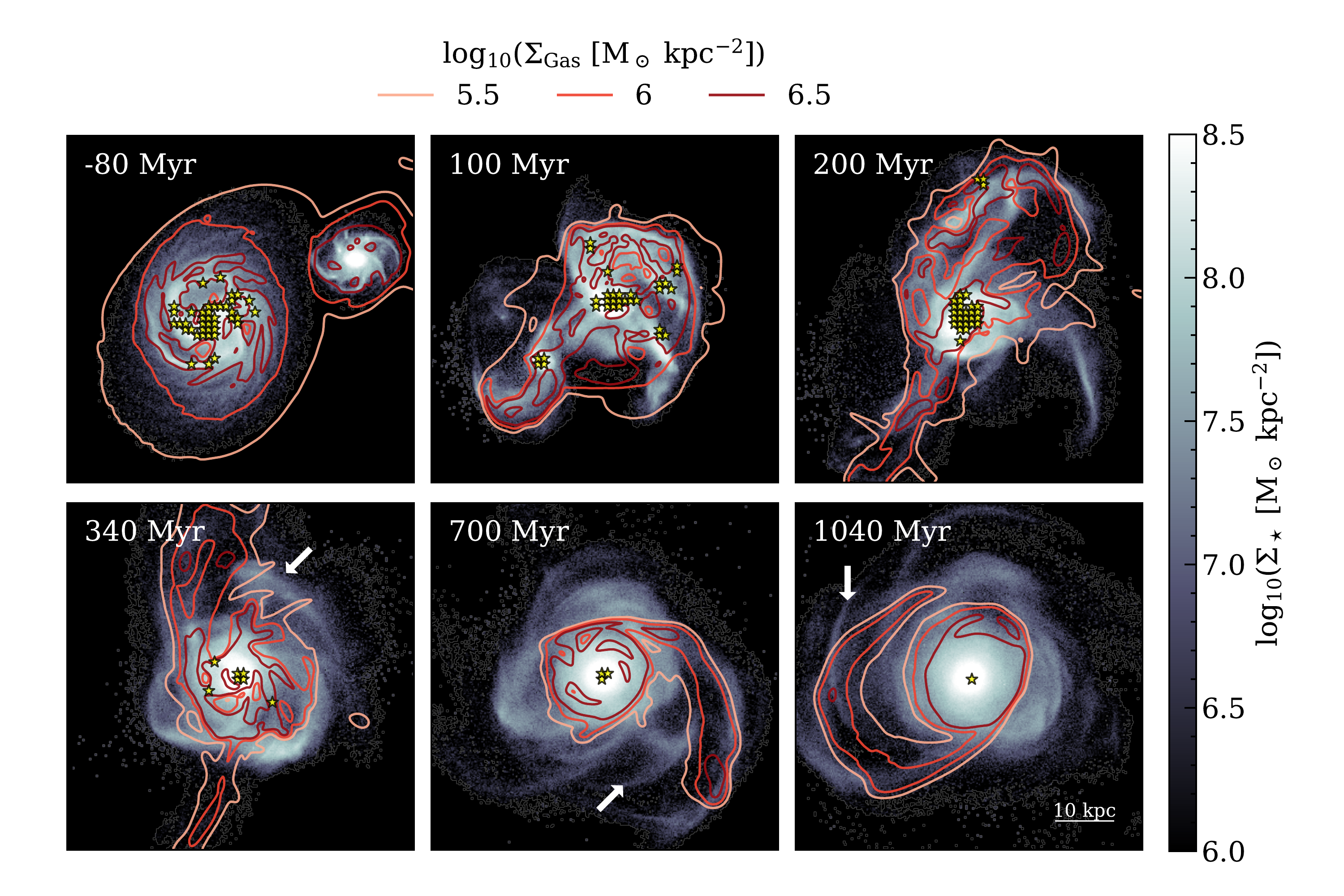}
	\caption{Face-on stellar surface density maps with contours of the gas surface density for the same snapshots as in Fig.~\ref{fig:composite}. Star-forming regions (traced by star younger then 10 Myr) (on a 1 kpc scale) are indicated with yellow stars (except for the satellite in the first panel at $t=-80$ Myr). Examples of stellar shells are indicated by white arrows.}
	\label{fig:shellsandgas}
\end{figure*}

\begin{figure*}
	\includegraphics[trim={1cm 1cm 1cm 1cm}, clip, width=2\columnwidth]{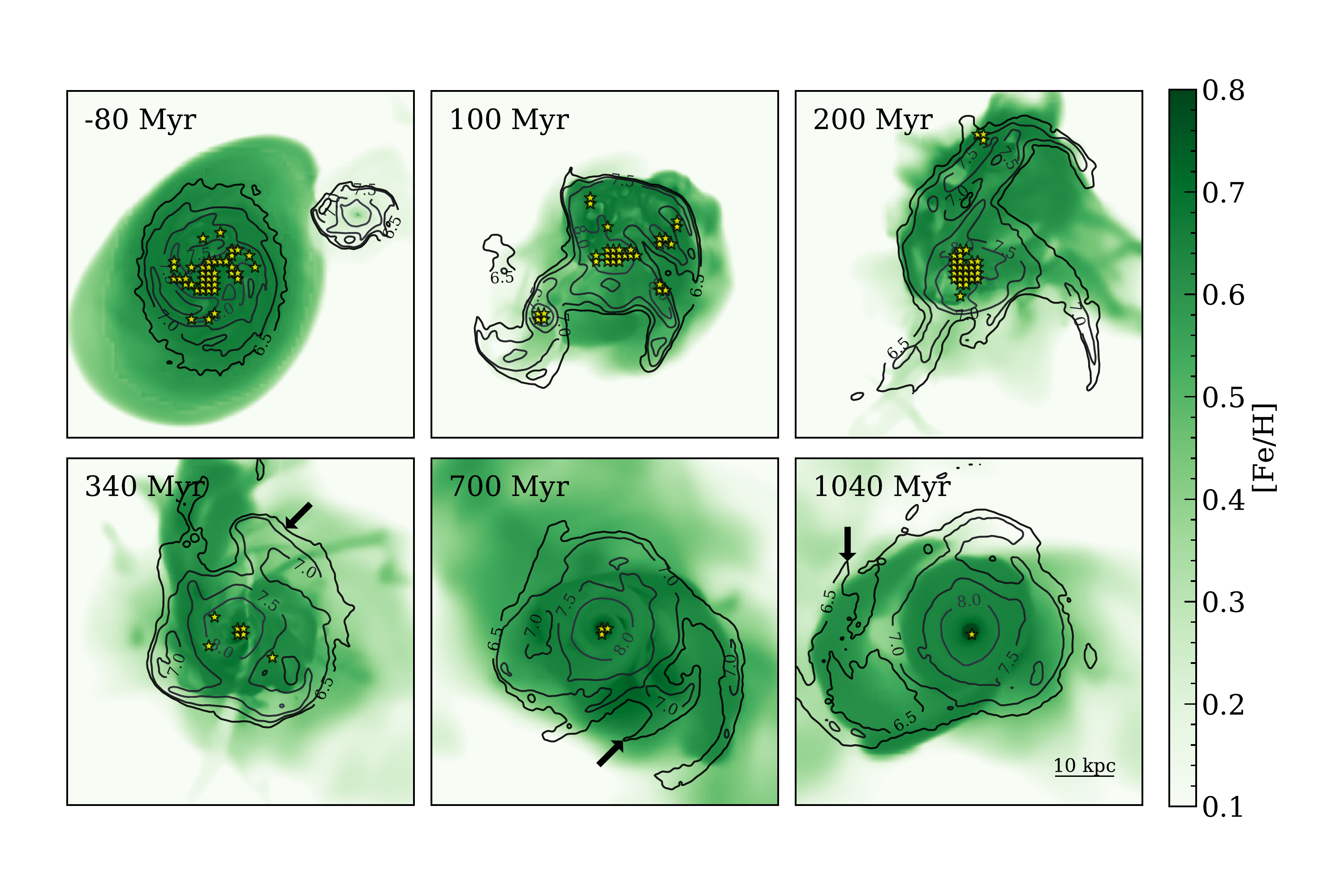}
	\caption{Face-on gas density-weighted maps of $\feh$ for the same snapshots as in Fig.~\ref{fig:composite}, with contours of the stellar surface density and yellow stars indicating where star formation occurs (similarly as in Fig.~\ref{fig:shellsandgas}). Note that even very diffuse gas is shown in these maps. The same examples of stellar shells as in Fig.~\ref{fig:shellsandgas} are marked with black arrows.}
	\label{fig:shellsandmetallicity}
\end{figure*}

\subsection{Shells and the distribution of gas}
\label{sect:shellsandgas}
In Section~\ref{sect:globalSFH} and~\ref{sect:resolvedSFA}, we show that the SFR drops because of morphological quenching for $t>250$ Myr, and that the off-centre star formation activity fades. Since the first shell does not appear until at $t \approx 340$ Myr, i.e. at a time when only a weak star formation activity remains and is exclusively confined in the galactic nucleus, no in situ star formation can occur in shells. Fig.~\ref{fig:shellsandgas} highlights the location of star-forming regions on maps of the stellar surface density with contours of the gas surface density. As expected, none of the three examples of stellar shells indicated with white arrows in the bottom panels show evidence of in situ star formation. All the shell structures are purely stellar, with no gaseous counterpart, in agreement with \citet{Weil1993}, and as expected for the outer shell of NGC 474 \citep{Mancillas2019a}. Indeed, the formation process of shells requires the conservation of orbital energy to allow the material of the satellite galaxy to pile up at its apocentres. While this conservation can be relatively well achieved in the discrete component that is the stellar one (with the caveat that dynamical friction progressively dissipates some of its orbital energy), the gas component is additionally subjected to shocks and ram pressure, specially on radial orbits. This slows the gas down considerably more than the stars, and prevents it from reaching the large galactocentric distances of the shells. Conversely, tidal tails mostly expand from the outskirts of the galactic discs \citep{Duc2013}, and thus experience much weaker and rarer shocks and ram pressure, such that the stellar and gaseous components remains largely coupled. It results that tidal tails, contrary to shells, have a prominent gas component which can possibly host in situ star formation, as it is the case in our simulation, e.g. in the upper part of the merger at $t=200$ Myr. This distinction between shells and tidal tails makes the two tidal features fundamentally distinct\footnote{By tracking the initial particles making up each galaxy, we find that the shells are mainly made up of stellar material from the satellite, while the tidal tails are instead mainly made up of material from the outer regions of the main galaxy.}, and suggests that shells can not host the formation of tidal dwarf galaxies.

Because of the absence of in situ star formation in the shell structures, it is tempting to use the metallicity of clusters detected there to backtrack them to their formation sites. However, this is made complicated by two aspects of the merger. First, the merger-induced starburst phase leads to a rapid enrichment of the ISM. In our simulation, the value of \feh in the central kpc increases by 0.5 dex in the 300 Myr following the first peak of SFR. Second, the accretion of the satellite implies a mixing of the respective gas reservoirs of the two galaxy progenitors, and a priori dilution of the metallicity of the main, as visible in Fig.~\ref{fig:shellsandmetallicity} (e.g. at 100 Myr and 200 Myr, where \feh decreases by $\sim$0.2--0.4 dex in various areas in the centre of the main galaxy). These two effects alter the chemical content of the ISM of the remnant over different timescales, varying with the intensity, compactness, and duration of the starburst(s), and the orbital configuration of the encounter. As a consequence, the time and spatial evolution of the metallicity in the remnant is very complex, and calls for caution when assessing the origin(s) of young massive star clusters (formed in the interaction) from their chemical content(s).

\section{Discussion}
\label{sect:discussion}
\subsection{Comparison with a real shell galaxy}
We find that a lack of molecular gas in shells prevents in situ star formation, and that the shell structures themselves form after the starburst phase induced by the merger. If our formation scenario applies to NGC 474, it suggests that the young massive star clusters detected in the outer shell of NGC 474 by \citet{Fensch2020} do not have an in situ origin. \citet{Fensch2020} proposes a formation scenario where the merger event responsible for the formation of the shells, also triggered the formation of massive star clusters via a nuclear starburst. Similarities in age and metallicity with the relatively young stellar population in the nucleus of NGC 474, indicates that this could indeed be the birth place of the young massive clusters. Their migration to the outer shell could be explained if they form from marginally stable gas structures at the moment when the satellite passes through the centre of the main galaxy. Assuming that the physical conditions for massive cluster formation would be met for long enough, and would overcome the destructive effects of tides and shear, the young massive clusters would then inherit the orbital energy of their parent gas structures, and would eventually continue their journey up to high galactocentric radii, i.e. up to the shells. However, it is possible that the coincidence of the observed clusters with the outer shell of NGC 474 is a projection effect. In such case, explaining the migration of clusters far from any other features would be more challenging, and would likely call for slingshots of the young massive clusters when they fly by the galactic centre. The limited resolution of our simulation does not allow for a detailed study on the formation of massive star clusters. However, we note that the onset of a starburst phase favors the formation of massive clusters \citep{Renaud2015}, which thus supports the scenario proposed by \citet{Fensch2020}.

\begin{figure*}
    \centering
    \includegraphics[width=2\columnwidth]{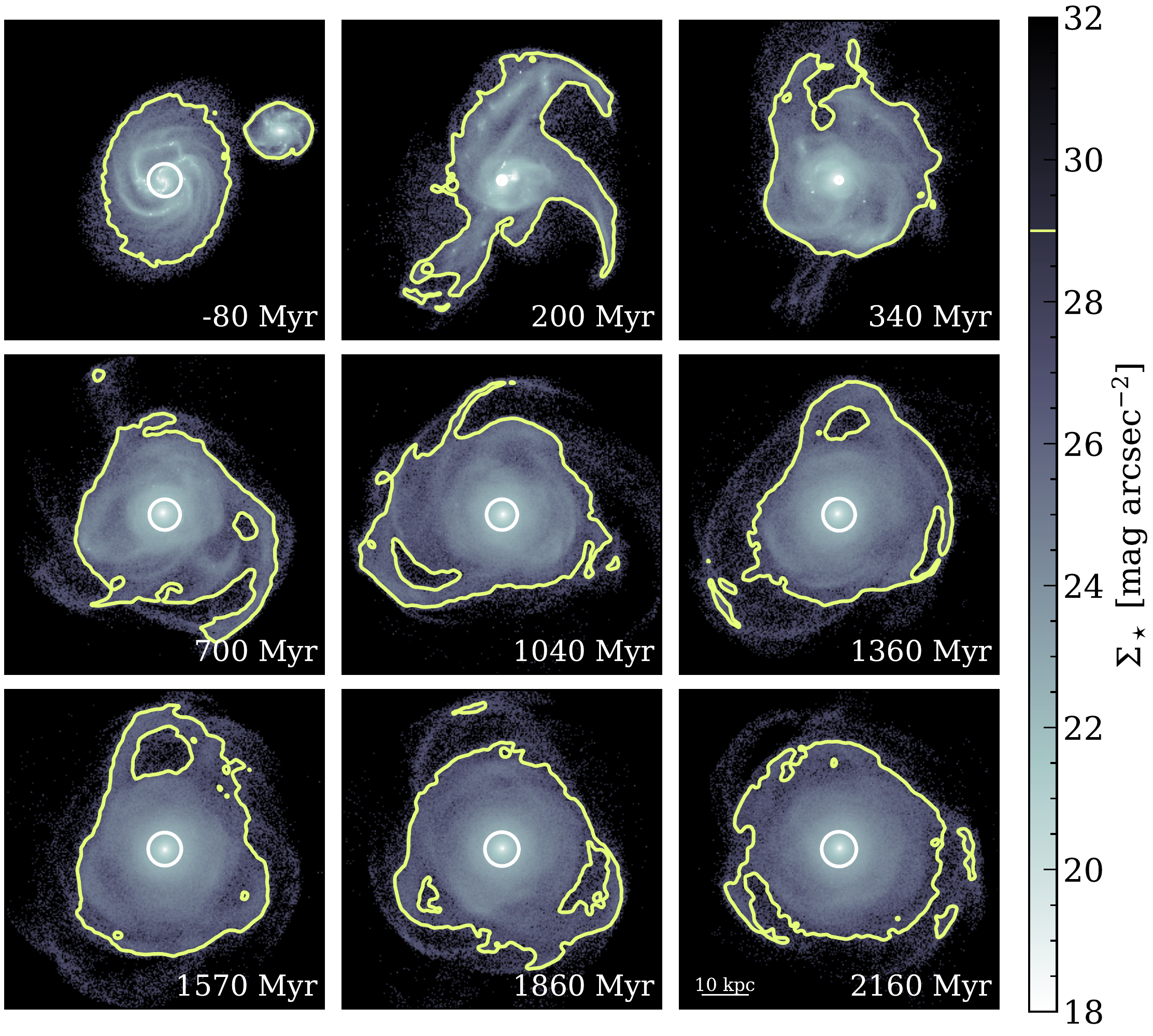}
    \caption{Mock images using the g-band filter of the MegaCAM instrument on the Canada-France-Hawaii Telescope (see Appendix~\ref{appendix:mock} for details). Snapshots with $t\gtrsim$ 340 Myr all include examples of stellar shells. The contours mark the surface brightness limit of MegaCAM, i.e. 29 mag arcsec$^{-2}$. White circles indicate the half-light radius of the system (but measured in the u-band).}
    \label{fig:mock}
\end{figure*}

An argument against the formation through a nuclear starburst in NGC 474 relies on the non-detection of H$_2$ in the nucleus \citep{Combes2007, Mancillas2019a}. In this case, the nuclear starburst must therefore have been very efficient at either consuming and/or ejecting all remaining molecular gas. However, the opposite is found in our simulation, where H$_2$ is still present in the central kpc after the starburst phase. An active galactic nucleus (AGN) fuelled by gravitational torques could efficiently remove molecular gas from the centre of the merger remnant, and even quench star formation \citep{Hopkins2013}. In our simulation however, quenching is reached globally, and without AGN feedback. Therefore, this leaves (small) amounts of molecular gas in the nucleus. We note that the two scenarios (with or without AGN) would have different signatures in the molecular gas contents, and can therefore serve as a tool to identify observationaly the cause of quenching in these types of mergers. In the case of NGC 474, we are not aware of any evidence of an AGN activity. Exploring further the nuclear starburst scenario calls for complementary observations of the nucleus of NGC 474, both of its molecular gas content and the presence of a AGN, combined with more advanced numerical simulations accounting for the possibility of ignition of an AGN. 

\begin{figure}
    \centering
    \includegraphics[width=\columnwidth]{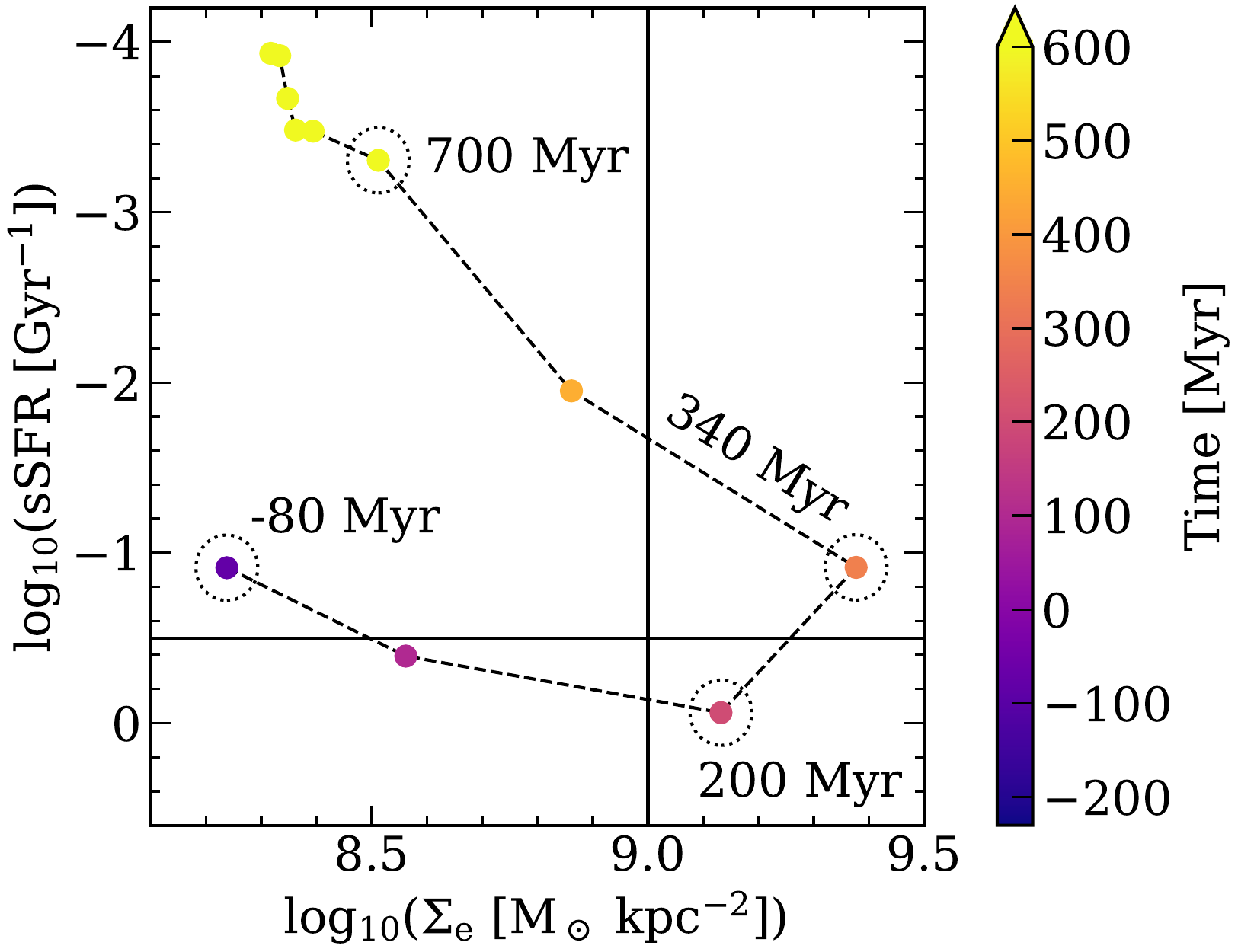}
    \caption{Time evolution of the sSFR versus the stellar surface density within two times the half-light radius of the main galaxy (measured in mock u-band maps). The time (with respect to the first pericentre passage) of four critical snapshots along the evolution on this plane are indicated with dotted circles. As in the compaction scenario of \citet{Dekel2014} and \citet{Zolotov2015}, the interaction transforms the system into a blue nugget ($t=200$ Myr). This is followed by a very short red nugget phase ($t=340$ Myr), and the system rapidly goes into a quenched non-compact state.}
    \label{fig:compaction}
\end{figure}

\subsection{Transformation into a red elliptical galaxy and the detectability of its faint shells}
\label{sect:shellsanderarlytypegalaxies}
In Section~\ref{sect:globalSFH}, we show that tidal interactions during the merger scatter stars into a stellar spheroid. What is left after the merger (besides tidal debris) is an elliptical galaxy (in Sect. \ref{sect:globalSFH} we show that $v_\mathrm{rot, s}/\sigma_\mathrm{s}$ decreases by a factor of $\sim$5) with a diffuse gas disc and very little to no star formation as a result of morphological quenching. \citet{Atkinson2013} finds that shells are more frequent around red early-type galaxies than blue late-types. One can therefore suspect that shell-forming mergers might be associated with the process of transforming blue late-type galaxies into red and dead early-types. If this is a generic feature of shell galaxies, the detection of shells around a galaxy could be used as an indicator of morphological quenching shutting down star formation, and due to the limited detectability of shells, possibly constrain the time of quenching \citep[see e.g.][]{Bilek2013, Ebrova2020}.

However, whether quenching is a generic feature of shell galaxies demands further investigation, and calls for more numerical simulations varying the orbital parameters and properties of the progenitor. This would provide valuable statistics on the effects the orbital configuration and mass ratio have on the quenching process. Furthermore, adding a cosmological context would provide insights into shell-forming mergers at different redshift, and whether a continuous accretion of gas could rejuvenate the merger remnant. 

Fig.~\ref{fig:mock} shows a series of mock surface brightness maps in the g-band of the MegaCAM instrument on the Canada-France-Hawaii Telescope, assuming a distance to the galaxy of 30.6 Mpc (see Appendix~\ref{appendix:mock} for details). Contours mark the surface brightness limit of 29 mag arcsec$^{-2}$ of MegaCAM. The early examples of shells in Fig.~\ref{fig:mock}, i.e. at 340 Myr, 700 Myr and 1040 Myr, are all detectable with this limit. However, some shells fall on its fainter side at later times. We note that real observations of shell galaxies yield more numerous and finer low surface brightness features than our mock images. The discrepancy could be caused by the limited resolution of the simulation, in term of sampling of the stellar component and softening of the gravitational potential, and/or the possible cosmological origin of these features (thus not captured in our non-cosmological setup).

Since the formation of shells specifically requires an almost radial encounter, shell-forming interactions should be a priori less frequent than those forming tidal tails. However, Fig.~\ref{fig:mock} suggests that shells are detectable over longer timescales than tails, where the latter dissolve and/or fade below the surface brightness limits within $\sim$1 Gyr \citep[see also][]{Mancillas2019a}. As a consequence, observations of shells around merger remnants could bias the statistics on the frequency of radial encounters.

\subsection{Compaction}
Our simulation shows several similarities with the compaction model, which explains the transformation of ``normal'' galaxies into compact star-forming systems (blue nuggets), followed by compact quenched spheroids (red nuggets) at high-redshift \citep{Dekel2014, Zolotov2015, Tacchella2016a, Tacchella2016b}. Figure~\ref{fig:mock} shows that the half-light radius of the main galaxy (measured using mock images in MegaCAM's u-band) shrinks during the interaction period (i.e. for a few 100 Myr), due to the enhanced central star formation activity. To examine this in more detail,  Fig.~\ref{fig:compaction} shows the time evolution of the specific star formation rate (sSFR) as a function of the stellar surface density\footnote{Following \citet{Zolotov2015}, the stellar surface density is defined as $\Sigma_\mathrm{e} = M_\mathrm{s}(r)/(\pi r^2)$, where $M_\mathrm{s}(r)$ is the enclosed stellar mass within a cylinder of radius $r$ equal to two times the half-light radius. The sSFR is defined as the total birth mass of the star particles younger than 10 Myr, divided by $M_\mathrm{s}(r)$, within the same cylinder.} within two times the half-light radius of the main galaxy. The black lines at $\log_{10}(\mathrm{sSFR} / \mathrm{Gyr}^{-1} ) = -0.5$ and $\log_{10}(\Sigma_\mathrm{e} / \mathrm{M}_\odot\, \mathrm{kpc}^{-2})  = 9.0$ mark the thresholds between star-forming versus quiescent, and compact versus non-compact galaxies, respectively \citep[as defined in][but for high-redshift, and thus smaller galaxies]{Zolotov2015}.

Our simulation shows the same first step to what is expected from a compaction event, i.e. a drastic loss of angular momentum (here, caused by tidal torques from the satellite) fuels the galactic centre with gas. This locally increases the gas density in the central $\sim$1 kpc and triggers a nuclear starburst. Despite non-negligible levels of off-centre star formation activity, this effectively shrinks the half-light radius (especially in bands tracing star formation), and forms a blue nugget. As such, we predict that the onset of a blue nugget should be a generic feature of disc galaxies involved in interactions, as long as the off-centre boost of star formation remains weak with respect to the nuclear activity \citep[see e.g.][on nuclear starbursts]{Keel1985, Moreno2021}.

In contrast to the cosmological simulations of \citet{Dekel2020}, the quenching process in our case is not due to the depletion of gas, but instead to morphological transformation (recall Section~\ref{sect:globalSFH}). The sudden stabilisation of the gas disc quenches the entire galaxy, meaning that the luminosity profile quickly flattens after that the nuclear starburst fades. The red nugget phase is therefore very short, and the galaxy moves to a quenched non-compact state (top left of Fig.~\ref{fig:compaction}) in a couple of 100 Myr. Because this transition is much faster than rejuvenation by gas accretion from a cosmological environment, we argue that this evolution is not affected by the non-cosmological nature of our simulation. 

\section{Conclusions}
\label{sect:conclusions}
We perform a $N$-body + hydrodynamical simulation of two disc galaxies and their near-radial major merger into a shell galaxy. We analyse the global star formation history and the spatially resolved star formation activity, to examine how the conditions of star formation activity evolve with time and within the system. Our main findings are as follows. 

\begin{enumerate}
    \item As the system approaches the first pericentre passage, an excess of dense gas is generated, likely by various physical processes, such as nuclear inflows, shocks, and tidal and turbulent compression. The star formation activity increases, and the system goes into a starburst phase with enhanced SFR and reduced depletion time. 
    \item Following the starburst phase at coalescence, the star formation activity transitions into a quenching phase, when a deficit of dense gas decreases the SFR and increases the depletion time of the system. The cause of the cessation of the star formation activity is morphological quenching: a stellar spheroid (formed in the merger by tidal scatter of stars) with its shallow gravitational potential, prevents the gas to collapse and form new stars. The star formation activity never recovers and the quenching phase continues throughout the rest of the simulation, for several Gyr. 
    \item The Kennicutt-Schmidt relation for star-forming regions on an 1 kpc-scale varies along the merger, but does not show any general trend with time. At low gas surface densities, the efficient mixing of HI and H$_2$ favoured by the interaction, generates a break in the relation, causing the power-law index of the star formation relation to be higher than the classical value of 1.4. The surface density of H$_2$ shows a better correlation to that of SFR than that of all the gas, while conversely, HI shows no correlation. This matches observations of the star formation relations in isolated disc galaxies. 
    \item Star formation occurs in regions with high molecular gas fraction, including the nucleus, spiral arms, and occasionally the outskirts of the system during the early stages of the merger. The quenching phase stops star formation without the need of AGN feedback, and leaves only the innermost $\sim$1.5 kpc with star-forming molecular gas, but at a SFR 1 dex lower than before the interaction. 
    \item The first shells appear after coalescence at large galactocentric distances when the off-centre star formation activity is quenched. In contrast to gas rich tidal tails formed in the early stages of the merger, shells are deprived of gas, and thus do not host the conditions necessary for in situ star formation. We propose that this is a generic feature of shell galaxies shaped by wet mergers, implying that the young massive star clusters detected in the outer shell of NGC 474 most likely did not form in situ.
    \item Shells keep a surface brightness above usual detectablity limits for longer than classical tidal tails. Therefore, the frequency of the specific type of encounters needed to form shells (i.e. near-radial orbits of a intermediate to major merger) could be over-estimated observationally if the timescale for their dissolution and fading is not correctly accounted for.
    \item Since the quenching process is closely linked to the morphological transformation of the system, our results suggest that shell-forming mergers might be part of the process in converting blue late-type galaxies into red and dead early-types.
	\item Our simulation shows similarities with the compaction scenario of \citet{Dekel2014, Zolotov2015, Tacchella2016a, Tacchella2016b}, where the interaction concentrates gas in the galactic centre and forms a blue nugget. Rather than depletion of gas, star formation ceases because of morphological quenching \citep{Martig2009}. The red nugget phase is therefore very short, as the galaxy rapidly transitions into a quenched non-compact state.
\end{enumerate}

\section*{Acknowledgements}
We thank the anonymous referee for their comments on how to improve this manuscript. FR and OA acknowledge support from the Knut and Alice Wallenberg Foundation, and from the Swedish Research Council (grant 2019-04659).
The simulations have been performed on the Tetralith supercomputer hosted at NSC (Link\"oping, Sweden), thanks to a SNIC allocation.

\section*{Data availability}
The data underlying this article will be shared on reasonable request to the corresponding author.



\bibliographystyle{mnras}
\bibliography{citations} 



\appendix

\section{Initial conditions}
The parameters for the galactic and orbital initial conditions are provided in Tables~\ref{tab:initial_conditions} and \ref{tab:orbital_parameters} respectively.

\begin{table}
	\centering
	\caption{Initial conditions for the galaxy models.}
	\label{tab:initial_conditions}
	\begin{tabular}{|lcc|} 
		\hline
		& Main & Satellite \\ 
		\hline
		\emph{Dark matter halo} (NFW) & & \\ 
		mass [$\times 10^{10} \Msun$] & 48.0 & 24.0 \\
		scale length [kpc] & 28.0 & 20.0 \\
		truncation radius [kpc] & 100.0 & 50.0 \\
		\hline
		\emph{Bulge} (Hernquist) & &\\
		mass [$\times 10^{10} \Msun$] & 0.5 & 0.5 \\
		scale radius [kpc] & 0.3 & 0.7 \\
		truncation radius [kpc] & 10.0 & 5.0 \\ 
		\hline
		\emph{Stellar disc} (exponential) & & \\
		mass [$\times 10^{10} \Msun$] & 3.5 & 1.5 \\
		scale radius [kpc] & 3.0 & 2.1 \\
		scale height [kpc] & 0.5 & 0.21 \\
		truncation radius [kpc] & 20.0 & 10.0 \\
		\hline
		\emph{Gas disc} (exponential) & & \\
		mass [$\times 10^{10} \Msun$] & 0.45 & 0.23 \\
		scale radius [kpc] & 4.0 & 2.8 \\
		scale height [kpc] & 0.3 & 0.28 \\
		truncation radius [kpc] & 30.0 & 15.0 \\
		\hline
	\end{tabular}
\end{table}

\begin{table}
	\centering
	\caption{Orbital initial conditions for the main and satellite galaxies. The non-zero value of $x$ in the spin axis of the main galaxy is what gives the satellite an initial inclination angle of $\sim 10\degree$ with respect to the plane of the main galaxy.}
	\label{tab:orbital_parameters}
	\begin{tabular}{|lcc|} 
		\hline
		& Main & Satellite \\ 
		\hline
		$(x, y, z)$ [kpc] & $(0, 0, 0)$ & $(60, 20, 0)$ \\
		$(v_x, v_y, v_z)$ [km s$^{-1}$]& $(0, 0, 0)$ & $(-30, 0, 0)$ \\
		Spin axis $(x, y, z)$ & $(0.1763, 0, 1)$ & $(0, 0.5, 0.5)$ \\
		\hline
	\end{tabular}
\end{table}

\section{Mock observations}
\label{appendix:mock}
To generate mock photometric maps, we follow the approach of \citet{Jonsson2006}, as briefly summarized below. We use tabulated spectral energy distributions of simple stellar populations from {\tt starburst99} \citep{Leitherer1999} to assign a spectrum to all the stellar particles, as a function of their age and metallicity. For the stars present in the initial conditions, we assign ages from a random distribution following a linearly decreasing star formation history for 13 Gyr, and we set their metallicity to the solar value. Our results are highly insensitive to these choices. For the stars formed during the simulation, we use their actual age and metallicities. The spectra are de-normalized using the mass of the stellar particles. Then, we multiply the spectra with the normalized transmission function of the selected filters. We assume a uniform dust-to-gas ratio of 1\% and use synthetic extinction curves from \citet{Draine2003}. We then compute the extinction for each particle along their line-of-sight using the three-dimensional distributions of the stars and the gas in the simulation. Finally, we integrate over wavelength and along the line-of-sight to obtained the flux in the corresponding band.

To convert the flux into surface brightness, we express it in AB absolute magnitude assuming a reference point of -48.6 mag, and then in apparent magnitude and surface brightness adopting a distance to the galaxy of 30.6 Mpc (corresponding to the distance to NGC~474, \citealt{Cantiello2007}).

\bsp	
\label{lastpage}
\end{document}